\documentclass[a4paper,11pt]{article}

\usepackage[utf8]{inputenc}
\usepackage[T1]{fontenc}
\usepackage[english]{babel}

\usepackage{indentfirst}
\usepackage{amsmath}
\usepackage{amsthm}
\usepackage{amssymb}
\usepackage{graphicx}
\usepackage{subfigure}
\usepackage{psfrag}
\usepackage{color}
\usepackage{pgf,pgfnodes}

\allowhyphens

\newcommand{\valos}{\mathbb{R}}
\newcommand{\complex}{\mathbb{C}}

\newcommand{\ordo}{\mathcal{O}}
\newcommand{\fa}{\mathfrak{a}}

\newcommand{\ket}[1]{{\left|#1\right\rangle}}
\newcommand{\bra}[1]{{\left\langle #1\right|}}

\newcommand{\vev}[1]{\left\langle #1 \right\rangle}


\makeatother

\begin{document}

\numberwithin{equation}{section}

\title{Generalized Gibbs Ensemble for Heisenberg Spin Chains}
\author{Bal\'azs Pozsgay$^1$\\
~\\
 $^{1}$MTA-BME \textquotedbl{}Momentum\textquotedbl{} Statistical
Field Theory Research Group\\
1111 Budapest, Budafoki út 8, Hungary
}
\maketitle

\abstract{
We consider the Generalized Gibbs Ensemble (GGE) in
 the context of global quantum quenches in XXZ Heisenberg spin
chains.
Embedding the GGE into the Quantum Transfer Matrix formalism we
develop an iterative procedure
to fix the Lagrange-multipliers
 and to calculate predictions
for the long-time limit of short-range correlators. 
The main idea is to consider truncated GGE's with only a finite number
of charges and to investigate the convergence of the numerical results
as the truncation level is increased.
As an example we consider a quantum
quench situation where the system is initially prepared in the N\'eel
state and then evolves with an XXZ Hamiltonian with anisotropy
$\Delta>1$. We provide predictions for short range correlators and
gather numerical evidence that the iterative procedure indeed converges. The
results show that the system retains memory of the initial condition,
and there are clear differences between the numerical values of the
correlators as calculated from the purely thermal and the Generalized Gibbs ensembles.
}

\section{Introduction}

Thermalization is a physical process whereby an isolated system reaches thermal
equilibrium through the interactions of its constituents.
Energy is conserved by time evolution, therefore the temperature
of the equilibrated system is determined solely by the available
energy at the beginning of the relaxation process. 

In quantum mechanics time evolution is represented by a unitary operation on
the Hilbert space. When at $t=0$ the system is in a state $\Psi_0$,
the density matrix 
evolves according to
\begin{equation*}
  \rho(t)=\ket{\Psi(t)}\bra{\Psi(t)}=
e^{-iHt}\ket{\Psi_0}\bra{\Psi_0}e^{iHt}
\end{equation*}
Therefore, the density matrix itself never thermalizes. Instead,
thermalization in quantum mechanics means that the expectation values
of physical observables (or equivalently the matrix elements of the
reduced density matrices) assume values which are equal to those
calculated in a thermal ensemble. 
In a generic quantum mechanical system this is expected to happen, if
there are no external driving forces 
present \cite{2008Natur.452..854R,Silva-quench-colloquium}.

The situation is different in the case of integrable models, in which
 there exists an infinite family of mutually
commuting quantum charges such that the Hamiltonian is a member of the
infinite series:
\begin{equation}
  \label{QM}
  \left[Q_j,Q_k\right]=0\quad\text{for}\quad j,k=1\dots\infty,
\qquad Q_2\sim H
\end{equation}
It follows that time evolution conserves the expectation
values of all charges, which prevents thermalization in the usual
sense. Instead, relaxation to a Generalized Gibbs Ensemble (GGE) was
proposed in \cite{rigol-gge}. The main idea is to construct a
thermodynamic ensemble where the statistical weights depend also on
all the higher charges and not only on the energy. It was proposed
that the long-time limit of quantum observables after a quantum quench
should be described by a GGE with appropriately chosen
Lagrange multipliers for the individual charges.

A huge amount of work was dedicated to study global quantum quenches in
integrable models (see
\cite{caux-essler-quench,Neel-quench-demler-gritsev} and references therein).
Most of the available results concern theories equivalent to free fermions 
\cite{mussardo-ising-1,mussardo-ising-2,ising-quench-1,ising-quench-2,ising-quench-3,ising-quench-4,ising-quench-5,essler-truncated-gge}
or interacting models in the CFT limit \cite{cardy-calabrese}. One
of the most important goals is to determine 
whether the GGE
provides a valid description of the possible steady states arising in
non-equilibrium situations. In the case of free theories the answer
seems to be a general affirmative yes
\cite{rigol2,ising-quench-2,essler-truncated-gge,gurarie-gge,second-quench,free-gge-1,free-gge-2,free-gge-3,free-gge-4}, 
whereas the question 
is still unresolved for
genuinely interacting theories. A further open problem in interacting
theories is how to use
the GGE to make actual predictions for 
physical observables. 

In Bethe Ansatz solvable models with only a single particle type 
the GGE hypothesis leads to a generalization of the Thermodynamic
Bethe Ansatz (TBA) framework  \cite{caux-GGE-LL,caux-GGE}. Originally
the TBA was developed to treat the purely thermal case \cite{YangYang2,Takahashi-book}, but the
addition of the higher charges does not modify its essential
properties \cite{caux-GGE}.
In principle the
generalized TBA (or GTBA) encodes all thermodynamic properties and it can give predictions for
the correlation functions as well \cite{davide,LM-sajat,marci-g3-megegy}. In the
Lieb-Liniger model studied in \cite{caux-GGE-LL,caux-GGE,caux-essler-quench},
however, this can not be carried out because the construction of
the higher charges seems to be an insurmountable problem
\cite{korepin-LL-higher} and therefore the Lagrange multipliers
entering the GGE can not be fixed. Instead, information about the
overlaps between pre-quench and post-quench states was used in
\cite{caux-GGE-LL} to set up the GGE.

It is a very natural idea to apply the GTBA formalism to XXZ spin
chains, where all higher charges can be constructed using their
definition via the transfer matrix, or alternatively with the help of
the boost operator
\cite{GM-higher-conserved-XXX,GM-higher-conserved-XXZ}. However, 
the infinite family of particle types (the strings of
different length) lead to
 an infinite set of GTBA equations. Although it has been
demonstrated 
that even such an infinite set can be treated numerically
\cite{caux-antoine}, fixing the Lagrange multipliers for 
the GGE through TBA seems extremely difficult.

An alternative to the TBA is the Quantum Transfer Matrix
(QTM) method \cite{kluemper-QTM,kluemper-review}, which leads to a
single non-linear integral equation 
replacing the infinite set of the TBA
\cite{TBA-QTM-Kluemper-Takahashi}. Moreover, there are very efficient
methods available to compute short-range correlations
with the QTM
\cite{XXZ-massless-corr-numerics-Goehmann-Kluemper,XXZ-massive-corr-numerics-Goehmann-Kluemper,XXZ-factorization-recent-osszefoglalo}. The
simplicity of the QTM with respect to the TBA makes it an excellent
candidate to describe the GGE. The idea of adding higher
charges to the QTM is not new: it was used in
\cite{Kluemper-GGE,kluemper-GGE2} to calculate thermal conductivities
and in \cite{kluemper-higher-1,kluemper-higher-2} to study the phase
diagram and thermal properties of integrable
spin chains with competing interactions.

In the present paper we apply the QTM formalism to set up the GGE for
the spin chains. As an application we consider a specific example of a global quench,
namely time evolution starting from the N\'eel state (in terms of the
anisotropy $\Delta$, this corresponds
to a quench from $\Delta=\infty$ to finite $\Delta$). Although the
relaxation of the antiferromagnetic order was already studied by
different methods in \cite{Neel-quench-1,Neel-quench-demler-gritsev},
these works did not consider the long-time limit of correlation functions.

The paper is organized as follows. In Section 2 we pose the problem in
general terms and explain our procedure to obtain the GGE as a limit of
 truncated GGE's. In Section 3 we present the extension of the QTM
 method to include a finite number of higher charges. Section 4
 includes calculations concerning the explicit form of the conserved
 charges and their mean values in the N\'eel state. In Section 5 we
 present our numerical results and finally we conclude in Section 6.

\section{Global quenches in integrable models}

Consider a 1D lattice model of $L$ sites with periodic boundary
conditions.
The local spin variables are $\sigma_j$ with $j=1\dots L$ and they can take $K$
values, therefore 
the Hilbert-space of the system is
\begin{equation*}
  \mathcal{H}=\otimes^L \left(\complex^K\right).
\end{equation*}
Consider a family of Hamiltonians of the form
\begin{equation}
\label{Hgen}
H=\sum_{j=1}^L  u_j+h_{j,j+1},
\end{equation}
where it is understood that $u_j$ and
$h_{j,j+1}$ are given by a 
translation of one-site and two site-operators $u_1$ and
$h_{1,2}$ which might depend on a finite set of coupling
constants. The generalization to Hamiltonians with multi-site 
interactions is straightforward. Note that in \eqref{Hgen} periodic
boundary conditions are assumed.

In this work we consider the situation of a sudden global quench. At $t=0$ we
prepare the system in a state $\ket{\Psi(t=0)}=\ket{\Psi_0}$ which
might be the ground state of a Hamiltonian $H_0$ or a state prepared
according to a well-defined rule. In the latter case we only require
that $\ket{\Psi_0}$ be defined in a natural way for any $L$ or at least any even $L$. 

At $t=0$ we change the parameters of the system such that
time evolution for $t>0$ will be governed by the
post-quench Hamiltonian $H$:
\begin{equation*}
  \ket{\Psi(t)}=e^{-iHt}\ket{\Psi_0}.
\end{equation*}

Consider a localized quantum observable $\mathcal{O}$. In the examples
to be investigated below $\mathcal{O}$ will be a correlation function
of local spin operators on neighboring sites or only a few sites
apart. The time dependence of this observable is
\begin{equation*}
  \ordo(L,t)=\bra{\Psi(t)}\ordo\ket{\Psi(t)}=
\bra{\Psi_0}e^{iHt} \ordo e^{-iHt}\ket{\Psi_0}.
\end{equation*}
In the formulas above it was understood that all quantities depend on the
volume $L$. We will be interested in the thermodynamic limit:
\begin{equation*}
\ordo(t)= \lim_{L\to \infty} \ordo(L,t).
\end{equation*}
Moreover we consider the infinite time limit of the observable:
\begin{equation}
  \bar\ordo=\lim_{t\to\infty} \ordo(t).
\end{equation}

\bigskip

In a generic non-integrable system we expect the phenomenon of thermalization. This
means that all long-time averages are described by a thermal
ensemble with an effective temperature $T=1/\beta$:
\begin{equation}
\label{thermal}
\bar\ordo=\frac{1}{Z} \text{Tr}\left( \ordo e^{-\beta H}\right)\qquad
Z=\text{Tr}\ e^{-\beta H}.
\end{equation}
Time evolution conserves the expectation value for the energy,
therefore the Lagrange multiplier $\beta$ can be fixed (at least in
principle) by the
requirement
\begin{equation}
\label{thermal2}
\vev{H}=-\frac{d}{d\beta}\log Z=  \bra{\Psi_0}H\ket{\Psi_0}.
\end{equation}
A central statement of thermalization is that there is a single
effective temperature $T$ for \textit{all} quantum observables.
Note that in formulas \eqref{thermal}-\eqref{thermal2} an implicit
infinite volume limit is understood. 

The situation is expected to be different when $H$ is
integrable, in which case the thermalization hypothesis 
does not hold. The reason for this is the following.
In integrable models there exists an infinite set of conserved
charges $Q_j$, $j=1\dots \infty$ which commute among themselves:
\begin{equation}
\label{con}
  [Q_j,Q_k]=0.
\end{equation}
The Hamiltonian is a member of the series, typically $H\sim
Q_2$ (the first charge $Q_1$ is usually the momentum operator). It
follows from \eqref{con} that the 
expectation values of all the charges are conserved in time. The $Q_j$
are constructed as sums of localized operators, therefore \eqref{thermal}
should apply if the system thermalizes.
However, the thermal ensemble would typically yield mean values for
the conserved charges which differ from those measured in the initial
state, therefore \eqref{thermal} can not be valid and thermalization can not occur. 

To describe the long-time average of observables in integrable models
the Generalized Gibbs Ensemble (GGE) was proposed in \cite{rigol-gge}. Setting aside
convergence properties for a moment the hypothesis of GGE can be
formulated as follows: There exists a set of couplings $\{\beta_j\}$
such that the long-time stationary state of the system is described by
the density matrix
\begin{equation}
\label{GGE0}
  \rho_{GGE}=\frac{1}{Z_{GGE}}\exp\left(-\sum_{j=1}^\infty \beta_j Q_j\right)\qquad
Z_{GGE}=\text{Tr}\exp\left(-\sum_{j=1}^\infty \beta_j Q_j\right).
\end{equation}
Physical quantities in this ensemble are given by
\begin{equation}
  \label{GGE1}
\ordo_{GGE}=\text{Tr}\left( \ordo \rho_{GGE}\right).
\end{equation}
Expectation values of the charges are conserved in time, therefore the
Lagrange multipliers can be fixed (at least in principle) by requiring
\begin{equation}
\label{GGE2}
\vev{Q_j}=-\frac{d}{d\beta_j}\log Z_{GGE}=  \bra{\Psi_0}Q_j\ket{\Psi_0}, \qquad
  j=1\dots\infty.
\end{equation}
The GGE hypothesis states that for any localized operator the
expectation value obtained by the equations \eqref{GGE0}-\eqref{GGE2}
is equal to the long-time average:
\begin{equation}
\label{senkisetudja}
  \ordo_{GGE}=\bar \ordo.
\end{equation}
Again, a crucial statement is that there is a single set of $\beta_j$
which determines \textit{all} physical observables. Note that all
 conserved charges need to be added to \eqref{GGE0}, otherwise the GGE
 can not be complete.

Equations \eqref{GGE0}-\eqref{GGE2} completely specify the GGE. However, in
practice it is impossible to fix all the 
$\beta_j$ and the convergence of the infinite sum is also not
guaranteed.  
In this work we propose to obtain the GGE as a limit of an iterative
procedure, where at step $k$ we only consider the first $k$
charges.  
This idea also appeared in the very recent paper
\cite{essler-truncated-gge} which concerned the quantum Ising chain.

To be specific, we define sets of parameters
$\{\beta^{(k)}_j,j=1\dots k\}$ which generate the density matrices of
truncated GGE's:
\begin{equation}
\label{GGEp}
  \rho^{(k)}=\frac{1}{Z^{(k)}}\exp\left(-\sum_{j=1}^k \beta_j^{(k)} Q_j\right)\qquad
Z^{(k)}=\text{Tr}\exp\left(-\sum_{j=1}^k \beta_j^{(k)} Q_j\right).
\end{equation}
We call the number $k$ the truncation level.
The $\beta_j^{(k)}$ are chosen such that
\begin{equation}
\label{GGEp2}
-\frac{d}{d\beta_j}\log Z^{(k)}=  \bra{\Psi_0}Q_j\ket{\Psi_0}, \qquad
  j=1\dots k.
\end{equation}
For the quantum observables we obtain a series
\begin{equation}
  \ordo^{(k)}=\text{Tr}\left( \ordo \rho^{(k)}\right).
\end{equation}
The actual GGE average is then given by the limit
\begin{equation}
\label{limo}
  \ordo_{GGE}=\lim_{k\to\infty} \ordo^{(k)}.
\end{equation}

This procedure provides a well-defined recipe to obtain the GGE
averages, but we are faced with the following questions:
Does the limit \eqref{limo} exist? Does it exist for all
 localized physical observables or maybe even non-local quantities like
  correlation lengths? How do the convergence properties depend on the
  Hamiltonian and the initial state?

While we can not answer these questions in their full generality, we
will present one non-trivial example in XXZ spin chains where
numerical evidence shows that the limit
\eqref{limo} exists for short-range correlation functions. This way the GGE
indeed gives predictions which could be compared to experiments or
independent numerical calculations.

\section{Thermal and Generalized Gibbs ensembles for the XXZ spin chain}

In this work we consider quantum quenches in spin-$1/2$ XXZ chains. The Hamiltonian 
is given by
\begin{equation}
  \label{XXZ-H}
  H_{XXZ}(J,\Delta,h)=J\sum_{j=1}^{L}
  (\sigma^x_j\sigma^x_{j+1}+\sigma^y_j\sigma^y_{j+1}+\Delta
(\sigma^z_j\sigma^z_{j+1}-1))+h\sum_{j=1}^L \sigma_j^z.
\end{equation}
In the following we will only consider the zero-field case
$h=0$. Moreover we set $J=1$ and restrict ourselves to the regime
$\Delta>0$. 

In the present section we set up the general framework of the GGE,
independent of the details of the initial 
state. Here we just assume that the mean values of the conserved charges can be
calculated with exact methods. Details of how to do this will be
provided in the next section, and we give a few comments 
about more general situations in the Conclusions.

The XXZ model is integrable for arbitrary $\Delta$. Its spectrum
is given by the Bethe Ansatz \cite{XXX,XXZ1,XXZ2,XXZ3} and the higher conserved charges can be
constructed using the transfer matrix formalism. 

Consider the R-matrix
acting on $\complex^2\otimes \complex^2$ given as
\begin{equation}
  R(u)=\frac{1}{\sinh(u+\eta)}
  \begin{pmatrix}
    \sinh(u+\eta) & & &\\
& \sinh(u)  & \sinh(\eta) & \\
& \sinh(\eta) & \sinh(u) & \\
& & & \sinh(u+\eta)
  \end{pmatrix}.
\label{R}
\end{equation}
Here $u$ is the spectral parameter and $\eta$ is a complex number
related to the anisotropy by $\Delta=\cosh\eta$.

The monodromy matrix is constructed as
\begin{equation*}
  T(u)=L_M(u)\dots L_1(u),
\end{equation*}
where $L_j(u)$ are local Lax-operators given by
\begin{equation*}
  L_j(u)=R_{0j}(u),
\end{equation*}
and the index 0  stands for the auxiliary spin space. In this space
$T(u)$ can be written as
\begin{equation*}
  T(u)=
  \begin{pmatrix}
    A(u) & B(u) \\
C(u) & D(u)
  \end{pmatrix},
\end{equation*}
where $A(u),B(u),C(u),D(u)$ are operators acting on the spin chain.

The transfer matrix is given by the trace
\begin{equation*}
  \tau(u)=\text{Tr}_0 T(u)=A(u)+D(u).
\end{equation*}

The $R$-matrix satisfies the Yang-Baxter equation \cite{Yang-nested,Baxter-book} which leads to the
commutativity of the transfer matrices:
\begin{equation*}
  [\tau(u),\tau(v)]=0.
\end{equation*}
This property is used to define the conserved charges of the model. It can be shown that
\begin{equation*}
  \tau(0)=U,
\end{equation*}
with $U$ being the translation operator on the chain  and it can be
considered as the first conserved charge: $U=e^{iQ_1}$, where $Q_1$ is
the momentum operator. The other charges are defined as logarithmic
derivatives of the transfer matrix at $u=0$:
\begin{equation}
\label{Qjdef}
  Q_j=\left(\frac{d}{du}\right)^{j-1} \log \tau(u).
\end{equation}
It was shown in \cite{Luscher-conserved} that the $Q_j$ defined this way are local
in the sense that they are given
as sums of products of spin variables such that they only span
a finite segment of the chain of length $j$. 
We note that using the normalizations above the second conserved
charge is 
\begin{equation*}
 Q_2=\frac{1}{2\sinh\eta} H_{XXZ}.
\end{equation*}

Eigenstates of the commuting family of transfer matrices can be
constructed using the Algebraic Bethe Ansatz \cite{korepinBook}. We choose the
reference state $\ket{F}$ to be the ferromagnetic state with all spins
up. Bethe states are then constructed by acting with the $B$-operators:
\begin{equation}
\label{BBB}
\ket{\mu_1,\dots,\mu_M}=\prod_{j=1}^M B(\mu_j) \ket{F}.
\end{equation}
Here the complex variables $\mu_j$ are the rapidities of the
interacting spin waves. Such a state is an eigenstate of the family of
commuting transfer matrices if the Bethe equations are satisfied:
\begin{equation}
  \label{BAe}
d(\mu_j)
\prod_{k\ne j}
\frac{\sinh(\mu_j-\mu_k+\eta)}{\sinh(\mu_j-\mu_k-\eta)}=1,
\end{equation}
where
\begin{equation}
\label{illuminus}
  d(u)=\bra{F}D(u)\ket{F}=  \left(\frac{\sinh(u)}{\sinh(u+\eta)}\right)^L.
\end{equation}
It can be shown that the states \eqref{BBB} are identical to the
states constructed by the coordinate Bethe Ansatz.

In the regime $\Delta>1$ we have
$\eta\in\valos$ and the usual one-string solutions of the Bethe equations
are of the form $\mu=i\lambda-\eta/2$ with $\lambda\in \valos$. On the
other hand, when $\Delta<1$ we have $\eta=i\gamma$ with
$\gamma\in\valos$ and the usual one-string solutions 
are $\mu=\lambda-i\gamma/2$ with $\lambda\in \valos$. The isotropic
case ($\Delta=1$)
can be obtained by a limiting procedure, but we do not consider this
 here.

In the following we will need the eigenvalues of the conserved charges
on the Bethe states. The Algebraic Bethe Ansatz yields the following
transfer matrix eigenvalue:
\begin{equation*}
t(u,\{\mu\}_M)=\prod_{j=1}^M \frac{\sinh(u-\mu_j-\eta)}{\sinh(u-\mu_j)}+d(u)
\prod_{j=1}^M\frac{\sinh(u-\mu_j+\eta)}{\sinh(u-\mu_j)}  .
\end{equation*}
Taking the logarithmic derivative gives
\begin{equation*}
  Q_j\ket{\mu_1,\dots,\mu_N}=
\left(\sum_{k=1}^N q_j(\mu_k)\right) \ket{\mu_1,\dots,\mu_N}
\end{equation*}
with 
\begin{equation}
q_j(\mu)=\left.\left(\left(\frac{\partial}{\partial u}\right)^{j-1} 
\log \frac{\sinh(u-\mu-\eta)}{\sinh(u-\mu)} \right)\right|_{u=0}.
\end{equation}
Here we assumed that $L>j$ such that the terms originating from $d(u)$
are all zero.

For the second charge (proportional to the one-string energy) we obtain
\begin{equation*}
  q_2(\mu)=\frac{\cosh(\mu)}{\sinh(\mu)}-\frac{\cosh(\mu+\eta)}{\sinh(\mu+\eta)}.
\end{equation*}
The formulas for the higher charges can be written in the form
\begin{equation}
\label{qj}
  q_j(\mu)=G_j(x_0)-G_j(x_+),
\end{equation}
where
\begin{equation*}
x_0=\frac{\cosh(\mu)}{\sinh(\mu)}\qquad\qquad
    x_+=\frac{\cosh(\mu+\eta)}{\sinh(\mu+\eta)}
\end{equation*}
and the $G_j(x)$ are polynomials satisfying the recursion
\begin{equation}
\label{Gj}
  G_{j+1}(x)=(x^2-1)\frac{d}{dx} G_j(x),\qquad G_2(x)=x.
\end{equation}

\bigskip

Thermal ensembles for the XXZ spin chains can be constructed using
the so-called Quantum Transfer Matrix (QTM) formalism
\cite{kluemper-QTM,kluemper-review}. The goal is to construct the
 density matrix
\begin{equation*}
  \rho=\frac{1}{Z} \exp(-\beta Q_2)\qquad
Z=\text{Tr}\exp(-\beta Q_2).
\end{equation*}
The central idea is to write down the Trotter-Suzuki
decomposition of $\exp(-\beta Q_2)$:
\begin{equation}
\label{Trotter-Suzuki}
  \exp(-\beta Q_2)\approx \left(1-\frac{\beta}{N} Q_2\right)^N 
\approx \left(\tau^{-1}(0) \tau(-\beta/N)\right)^N.
\end{equation}
Here $N$ is called the Trotter number and the two relations above
become equalities in the $N\to \infty$ limit. 
At finite $N$ the trace
\begin{equation*}
  Z_{N,L}=\text{Tr} \left(\tau^{-1}(0) \tau(-\beta/N)\right)^N
\end{equation*}
can be interpreted as a partition function of a six-vertex model with
$L$ vertical and $N$ horizontal lines. The vertical lines correspond
to the original spin spaces and the horizontal ones are auxiliary
spaces. The partition function can be obtained alternatively by ''quantizing'' the
system in the horizontal direction introducing the ''Quantum Transfer
Matrix'' which acts on the auxiliary spaces. The $L\to\infty$ limit of
the partition function can then be obtained by the largest eigenvalue
of the QTM. It is known that the QTM is gapped in the
sense that the second largest eigenvalue is separated from the largest
by a finite distance even in the $N\to\infty$ limit. This means that
all thermodynamic properties will be determined by the single leading eigenstate of
the QTM. This state can be found by the Algebraic Bethe Ansatz and its
transfer matrix eigenvalues can be computed both at finite $N$ and in
the Trotter limit. For the details of this procedure we refer the
reader to the review \cite{kluemper-review}. 

In the Trotter limit the partition function can be expressed as
\begin{equation*}
\log Z=-fL+\dots,
\end{equation*}
where the dots denote exponentially small corrections in $L$ and the free
energy density is given by
\begin{equation}
  f=- \int_C \frac{d\omega}{2\pi i} 
\frac{\sinh\eta \log(1+\fa(\omega))}{\sinh\omega\sinh(\omega+\eta)}.
\end{equation}
Here $\fa(\lambda)$ is an auxiliary function defined on the complex
plane which satisfies the nonlinear integral equation (NLIE)
\begin{equation}
\label{NLIE0}
\begin{split}
\log \fa(\lambda)=&
-\beta q_2(\lambda)
-  \int_C \frac{d\omega}{2\pi i} 
\frac{\sinh2\eta \log(1+\fa(\omega))}{\sinh(\lambda-\omega+\eta)\sinh(\lambda-\omega-\eta)}.
\end{split}
\end{equation}
The contour $C$ in the equations above depends on $\Delta$. For the
$\Delta>1$ regime considered in the present work it can be chosen as a
union of two straight line segments:
\begin{equation*}
C= [-i\pi/2+\alpha,i\pi/2-\alpha]\cup
 [i\pi/2-\alpha,-i\pi/2+\alpha],
\end{equation*}
where $\alpha<\eta/2$ is an arbitrary parameter\footnote{When a
  magnetic field is added, $\alpha$ has to be large enough such that
  the contour encircles all Bethe roots of the QTM.}.
Note that the first line segment runs upwards and the second runs downwards.

Thermodynamic properties of the spin chain can be calculated by taking derivatives of
the free energy with respect to the physical parameters.
Correlation functions are also accessible to this method.
In \cite{QTM1} multiple
integral formulas were calculated for the localized correlation functions. 
These were later found to factorize,  ie. they can be expressed as sums of products of simple
integrals \cite{XXZ-finite-T-factorization} so that numerical results
 can be produced in a very efficient way \cite{XXZ-massless-corr-numerics-Goehmann-Kluemper,XXZ-massive-corr-numerics-Goehmann-Kluemper,XXZ-factorization-recent-osszefoglalo}.

\bigskip

It is a very natural idea to use the Trotter-Suzuki decomposition to
construct the truncated GGE density matrices \eqref{GGEp}.
In fact this method was already worked out in \cite{Kluemper-GGE} where it
was used to obtain thermal conductivities. 

The main idea is that for any finite $k$ the decomposition
\begin{equation}
  \label{TS2}
  \exp(-\sum_{j=2}^k \beta_j Q_j)\approx \left(1-\frac{\sum_{j=2}^k \beta_j Q_j}{N}+o(1/N) \right)^N 
\end{equation}
holds. The right hand side can be obtained as
\begin{equation}
1-\frac{\sum_{j=2}^k \beta_j Q_j}{N}=
\left(\tau^{-1}(0)\right)^{k-1} \tau(u_1)\tau(u_2)\dots \tau(u_{k-1})+o(1/N),
\end{equation}
with some appropriately chosen numbers $u_j$. For example in the
 case of $k=3$ and non-zero $\beta_3$ a solution is
\begin{equation*}
  u_1=\sqrt{\frac{\beta_3}{N}}-\frac{\beta_2}{2N}\qquad
 u_2=-\sqrt{\frac{\beta_3}{N}}-\frac{\beta_2}{2N}.
\end{equation*}
Assuming that the numbers $u_j$ are found the partition function can
be expressed as
\begin{equation}
 Z_{N,L,k}=\text{Tr}\left[ 
\left(\tau^{-1}(0)\right)^{k-1} \tau(u_1)\tau(u_2)\dots \tau(u_{k-1})\right]^N.
\end{equation}
This is equivalent to a partition function of a six-vertex
model, which can be quantized in the horizontal direction, leading to a
modification of the original thermal QTM with a different set of 
inhomogeneities. We can assume that the analyticity properties
necessary for the construction of the Trotter limit hold also in the
modified problem, at least in a small neighborhood of the purely thermal
case. 
Then the Trotter limit can be taken and it leads to
\begin{equation}
\log   Z_{L,k}= -f^{(k)}L+\dots
\end{equation}
with
\begin{equation}
\label{fk}
  f^{(k)}=- \int_C \frac{d\omega}{2\pi i} 
\frac{\sinh\eta \log(1+\fa^{(k)}(\omega))}{\sinh\omega\sinh(\omega+\eta)}.
\end{equation}
Here $\fa^{(k)}$ is the auxiliary function solving the modified NLIE
\begin{equation}
\label{NLIE1}
\begin{split}
\log \fa^{(k)}(\lambda)=&
-\sum_{j=2}^k\beta_j q_j(\lambda)
-  \int_C \frac{d\omega}{2\pi i} 
\frac{\sinh2\eta \log(1+\fa^{(k)}(\omega))}{\sinh(\lambda-\omega+\eta)\sinh(\lambda-\omega-\eta)}.
\end{split}
\end{equation}
Note that the structure of the source term reflects the form of the
truncated GGE density matrix \eqref{TS2}.

Expectation values of conserved charges in the truncated GGE can be
obtained by
\begin{equation}
  \vev{Q_j}=L\frac{d}{d\beta_j} f^{(k)}.
\end{equation}
Instead of taking the derivative of the formula \eqref{fk} it is
useful to express $f^{(k)}$ with the 
functions
$\bar\fa^{(k)}=1/\fa^{(k)}$. It can be shown that they satisfy
 \begin{equation}
\label{NLIE2}
\begin{split}
\log \bar\fa^{(k)}(\lambda)=&
-\sum_{j=2}^k\beta_j q_j^-(\lambda)
+  \int_C \frac{d\omega}{2\pi i} 
\frac{\sinh2\eta \log(1+\bar\fa^{(k)}(\omega))}
{\sinh(\lambda-\omega+\eta)\sinh(\lambda-\omega-\eta)}
\end{split}
\end{equation}
with
\begin{equation}
\label{qjm}
  q_j^-(\mu)=G_j(x_0)-G_j(x_-),
\end{equation}
where
\begin{equation*}
x_0=\frac{\cosh(\mu)}{\sinh(\mu)}\qquad\qquad
    x_-=\frac{\cosh(\mu-\eta)}{\sinh(\mu-\eta)}.
\end{equation*}
The polynomials $G_j$ are defined by \eqref{Gj}. 
The free energy can be expressed with $\bar\fa^{(k)}$ as
\begin{equation}
\label{fk2}
  f^{(k)}=\int_C \frac{d\omega}{2\pi i} 
\frac{\sinh\eta \log(1+\bar\fa^{(k)}(\omega))}{\sinh\omega\sinh(\omega-\eta)}.
\end{equation}
Taking the derivative of \eqref{NLIE2} and \eqref{fk2} with respect to $\beta_j$ we
introduce the functions 
\begin{equation*}
  \fa_{j}^{(k)}(\lambda)=-\frac{1}{\fa^{(k)}(\lambda)}\frac{\partial
    \fa^{(k)}(\lambda)}{\partial \beta_j}=
\frac{1}{\bar\fa^{(k)}(\lambda)}\frac{\partial
    \bar\fa^{(k)}(\lambda)}{\partial \beta_j}.
\end{equation*}
They satisfy the linear equations
\begin{equation}
\label{NLIEmderiv}
\begin{split}
\fa_{j}^{(k)}(\lambda)=&
- q_{j}^-(\lambda)
+ \int_C \frac{d\omega}{2\pi i} 
\frac{\sinh2\eta
}{\sinh(\lambda-\omega+\eta)\sinh(\lambda-\omega-\eta)}
\frac{\fa_j^{(k)}(\omega)}{1+\fa^{(k)}(\omega)}.
\end{split}
\end{equation}
Finally the conserved charges in the truncated
GGE are given by
\begin{equation}
\label{Qj}
  \vev{Q_j}=L
 \int_C \frac{d\omega}{2\pi i} 
\frac{\sinh\eta }{\sinh\omega\sinh(\omega-\eta)}
\frac{\fa_j^{(k)}(\omega)}{1+\fa^{(k)}(\omega)}.
\end{equation}

With this we have finished the construction of the truncated GGE using
the QTM formalism. The remaining task is to calculate
correlation functions in the truncated GGE. 

In the derivation of the multiple integrals of \cite{QTM1} for the thermal
correlations it is not necessary to know the exact position of the
Bethe roots of the QTM; the only required information is that they can
be surrounded by appropriate contours and that there exists an
auxiliary function $\fa(\lambda)$ which encodes the positions of the
Bethe roots through the equation $\fa(\lambda)=-1$ and which has a
well-behaving Trotter limit. These conditions also hold for the
truncated GGE's,  at least in the neighborhood of a
thermal case with a finite $\beta_2$.
In the multiple integrals of \cite{QTM1}
the auxiliary function (or the 
combination $1/(1+\fa(\lambda))$ plays the role of a weight function,
therefore the representation of \cite{QTM1} is valid also in the truncated
GGE, provided that zeroes of $(1+\fa(\lambda))$ do not
cross the contours.
Based on
continuity and symmetry arguments we expect 
that the Bethe roots
will still be situated on the imaginary axis (when $\Delta>1$) or the
real axis (when $\Delta<1$). Our
numerical findings support this expectation in all cases we
encountered (see Section 5).  We conclude that the multiple integrals
of \cite{QTM1} are also valid in the truncated GGE's, at least in
the cases considered in the present work.

Factorized formulas for the correlators were developed in \cite{XXZ-finite-T-factorization}.
This factorization of the multiple integrals depends only on certain
algebraic properties underlying the construction of 
correlation functions and not on the particular physical parameters
(finite temperature or finite size)
of the problem at hand 
\cite{hidden-fermionic-structure-review,XXZ-factorization-recent-osszefoglalo}. It
follows that
all the formulas 
already available for the thermal case
are also valid in
the truncated GGE, provided the calculations are carried out using the
auxiliary functions $\fa^{(k)}(\lambda)$ which satisfy
\eqref{NLIE1} instead of \eqref{NLIE0}. We
refrain here from replicating the necessary formulas and 
refer the reader to the original papers
\cite{XXZ-finite-T-factorization,XXZ-factorization-recent-osszefoglalo}.

The method we described can be applied to construct the GGE within the QTM
formalism and to calculate correlation functions in the GGE. The
only input from the initial states is through the expectation values
\begin{equation*}
  \vev{Q_j}=\bra{\Psi_0}Q_j\ket{\Psi_0}
\end{equation*}
which are used to fix the Lagrange multipliers. These mean values
should be obtained exactly with use of explicit formulas for the $Q_j$
or by other methods. In the next section we show one example
where this task can be performed relatively easily.

\section{Quench dynamics from the N\'eel state}

We consider a quench  with the N\'eel state as initial state:
\begin{equation*}
\ket{\Psi_0}=\ket{N}=\ket{+-+-+-\dots}.
\end{equation*}
This state is not translationally invariant,
 the correlation functions
involving an odd number of spin operators (such as the magnetization
$\vev{\sigma_j^z}$ itself) have an obvious position dependence. However, we expect that in the
long time limit translational invariance of all correlation functions
will be restored \cite{Neel-quench-1}, in particular
\begin{equation*}
  \lim_{t\to\infty} \vev{\sigma_j^z}=0.
\end{equation*}
We note that even though the magnetization relaxes to zero,
traces of the original antiferromagnetic order are expected to show up in
the long-time limit of two-point correlators $\sigma_{j}^z\sigma_{j+1}^z$ and
$\sigma_{j}^x\sigma_{j+1}^x$. Our numerical result show that this is
indeed the case.

As an alternative to the N\'eel state we could also choose the initial states
\begin{equation*}
\ket{\Psi_0}_{\pm}=\frac{1}{\sqrt{2}}(\ket{N}\pm \ket{AN}),
\end{equation*}
where $\ket{AN}=\ket{-+-+-+\dots}$ is the
anti-N\'eel state. These states
 lead to correlation
 functions which are translationally invariant at all times.
Moreover, the expectation values of the conserved charges are equal to the
respective values in the N\'eel state, 
 because in the infinite volume
case considered here all cross-terms of the
form $\bra{N}Q_j\ket{AN}$ vanish. 
We note that the cross terms do not influence the time-dependent correlations
either, because
\begin{equation}
\label{crosss}
  \bra{N}e^{iHt}\ordo e^{-iHt}\ket{AN}=0
\end{equation}
at any finite $t$, given that $\ordo$ is a localized operator
and the infinite volume limit is already performed\footnote{In finite
  volume there are non-zero contributions to \eqref{crosss} when the
  $t$ is large enough so that quasiparticles can travel around the
  volume and therefore a complete shift of the two states is
  possible.}. Therefore the predictions of the same GGE hold in all three
cases, but it is enough to calculate the conserved charges in $\ket{N}$.

The N\'eel state is not an eigenstate of the Hamiltonian, therefore
the transfer matrix itself can not be used to calculate the
mean values directly and the explicit form of the $Q_j$ is
required. 

\subsection{Conserved charges in the N\'eel state}

Higher conserved charges of XXX and XYZ spin chains were considered in
the papers \cite{GM-higher-conserved-XXX,GM-higher-conserved-XXZ}. The
basic tool of these papers is the boost operator
\cite{Tetelman,Thacker-boost} given (in the XXZ case) by the formal
expression 
\begin{equation*}
B=\sum_{j=-\infty}^{\infty}
j  (\sigma^x_j\sigma^x_{j+1}+\sigma^y_j\sigma^y_{j+1}+\Delta
\sigma^z_j\sigma^z_{j+1})  .
\end{equation*}
It can be shown that the boost operator generates the derivative of
the transfer matrix with respect to the spectral parameter:
\begin{equation}
\label{booo}
  [B,T(u)]\sim \dot T(u).
\end{equation}
It follows that conserved charges are generated recursively as
\begin{equation*}
  [B,Q_j]\sim Q_{j+1}.
\end{equation*}
The equation above is to be understood up to additive constants.

The authors of  \cite{GM-higher-conserved-XXX,GM-higher-conserved-XXZ}
considered the action of the commutator \eqref{boo} 
 and derived a
recursive procedure to obtain all terms of the charges. They were able
to analytically solve the recursion in the XXX case, whereas in the
general XYZ case (including the XXZ chains) they derived the explicit
form of the charges up to $Q_6$.

In order to translate the results of
\cite{GM-higher-conserved-XXX,GM-higher-conserved-XXZ} into our
conventions we define operators $Q_{j}^{\text{GM}}$ as
\begin{equation}
\label{boo}
 \frac{1}{2} [B,Q_j^{\text{GM}}]= Q_{j+1}^{\text{GM}}
\end{equation}
with the first member being
\begin{equation*}
Q_{2}^{\text{GM}}=\sum_{j=-\infty}^{\infty}
  (\sigma^x_j\sigma^x_{j+1}+\sigma^y_j\sigma^y_{j+1}+\Delta
\sigma^z_j\sigma^z_{j+1})  .
\end{equation*}
Up to overall phase factors these operators coincide with the ones given in 
\cite{GM-higher-conserved-XXX,GM-higher-conserved-XXZ}
and they differ from the $Q_j$ used in the
present work in additive and multiplicative normalization. The
additive constants can be fixed by requiring that all mean values
vanish in the reference state $\ket{F}$; this follows from the fact
that in our normalization
\begin{equation*}
  \bra{F}\tau(u)\ket{F}=1.
\end{equation*}
The multiplicative normalization can be deduced by working out the
proof of \eqref{booo} presented in \cite{GM-higher-conserved-XXZ}
using our normalizations.
We find the relation
\begin{equation}
\label{GMitt}
  Q_j=\frac{Q_j^{\text{GM}}-\bra{F}Q_j^{\text{GM}}\ket{F}}{2 (\sinh\eta)^{j-1}}.
\end{equation}

An important basic result of
\cite{GM-higher-conserved-XXZ} for the XYZ chain is that 
each conserved charge $Q_j$ is a sum of terms of the form
\begin{equation}
\label{opp}
[(\dots (  \hat\sigma_{i_1}\times \tilde\sigma_{i_2})\times
  \tilde\sigma_{i_3})\dots \times \tilde \sigma_{i_{l-1}}]
 \cdot \hat\sigma_{i_l}.
\end{equation}
Here $(i_1,i_2,\dots,i_l)$ is a sequence of sites in increasing
order such that $l\le j$ and $i_l-i_1<j$. The notation $ \hat\sigma$
and $\tilde\sigma$ refers to vectors constructed out of rescaled
Pauli-matrices. In the XXZ case 
they are
\begin{equation*}
\hat\sigma=(\sigma^x,\sigma^y,\sqrt{\Delta}\sigma_z)\qquad\qquad
\tilde\sigma=(\sqrt{\Delta}\sigma^x,\sqrt{\Delta}\sigma^y,\sigma^z).
\end{equation*}
In the N\'eel state only those terms are
non-vanishing which only include $\sigma_z$ operators. From all
possible terms in \eqref{opp} we find such products 
only when $l=2$, because any cross-product would necessarily involve at
least one $\sigma_x$ or $\sigma_y$ operator. Therefore it is enough to
collect the terms of the form
$\sigma^z_i\sigma_{i+l}^z$.
These terms only appear in the even charges $Q_{2j}$, which is
consistent with the general statement that the mean values of the
odd charges vanish in any parity invariant state.

The paper \cite{GM-higher-conserved-XXZ} provides formulas up to
$Q_6$. In order to obtain explicit representations for the higher
charges we implemented the iteration procedure \eqref{boo} with the
symbolic manipulation program \texttt{form} \cite{Vermaseren}. 
The formal commutation relation \eqref{boo} is only valid on an
infinite spin chain, but it can still be used to calculate the
charges.
Defining the finite size operators
\begin{equation*}
B^{l}=\sum_{j=1}^{l-1}
j  (\sigma^x_j\sigma^x_{j+1}+\sigma^y_j\sigma^y_{j+1}+\Delta
\sigma^z_j\sigma^z_{j+1})  
\end{equation*}
\begin{equation*}
Q_{2}^{l}=  (\sigma^x_l\sigma^x_{1}+\sigma^y_l\sigma^y_{1}+\Delta
\sigma^z_l\sigma^z_{1})  +\sum_{j=1}^{l-1}
  (\sigma^x_j\sigma^x_{j+1}+\sigma^y_j\sigma^y_{j+1}+\Delta
\sigma^z_j\sigma^z_{j+1})  
\end{equation*}
we implemented the recursion
\begin{equation}
\label{boo2}
 \frac{1}{2} [B^l,Q_j^{l}]= Q_{j+1}^{l}.
\end{equation}
At each step of the iteration additional boundary terms arise which
start to propagate towards the middle of the chain. However, if $l$ is
large enough then the ''bulk'' of $Q_{j}^l$ is not affected and
the coefficients of the different terms of $Q_{j}^{\text{GM}}$ can be read off from the
middle of the chain. We used this method to obtain all charges up to
$Q_{12}$. For the sake of brevity we only present the terms
relevant for the N\'eel state, and only up to $Q_8$:
\begin{equation}
  \label{GM}
  \begin{split}
 Q_2^{GM} &=\sum_j \Delta \sigma_j^z\sigma_{j+1}^z+\dots\\
   Q_4^{GM} &=\sum_j    4\Delta \sigma_j^z\sigma_{j+1}^z- 2\Delta \sigma_j^z\sigma_{j+2}^z+\dots\\
   Q_6^{GM} &=\sum_j  (56\Delta + 16\Delta^3) \sigma_j^z\sigma_{j+1}^z -( 64\Delta + 8\Delta^3)\sigma_j^z\sigma_{j+2}^z + 24\Delta\sigma_j^z\sigma_{j+3}^z+\dots\\
   Q_8^{GM} &= \sum_j  (1504\Delta + 1312\Delta^3 +
   64\Delta^5)\sigma_j^z\sigma_{j+1}^z - (2912\Delta + 1376\Delta^3 +
   32\Delta^5)\sigma_j^z\sigma_{j+2}^z +\\ &\hspace{1cm}+ (2400\Delta
   + 480\Delta^3)\sigma_j^z\sigma_{j+3}^z -
   720\Delta\sigma_j^z\sigma_{j+4}^z+\dots
\end{split}
\end{equation}
The dots represent terms with at least two $\sigma^x$ or $\sigma^y$ operators.

The N\'eel state mean values read (up to $Q_{12}$)
\begin{equation}
\label{NN}
\begin{split}
Q^N_2&=-2\Delta      \\   
Q^N_4&=-8\Delta      \\   
Q^N_6&=-(160\Delta+32\Delta^3    )  \\   
Q^N_8&=- (7808\Delta +3584\Delta^3 +128\Delta^5    ) \\   
Q^N_{10}&=-(  709120\Delta + 517632\Delta^3 +62976\Delta^5 + 512\Delta^7)\\
Q^N_{12}&=- ( 103467008\Delta + 103763968\Delta^3 +23973888\Delta^5 + 1036288\Delta^7 +2048\Delta^9),
\end{split}
\end{equation}
where we defined
$Q_j^N=(\bra{N}Q_j^{\text{GM}}\ket{N}-\bra{F}Q_j^{\text{GM}}\ket{F})/L$. 

For the sake of completeness we also present the ferromagnetic
expectation values of $Q_{j}^{\text{GM}}$. Defining $Q_j^F=\bra{F}Q_j^{\text{GM}}\ket{F}/L$
we obtain
\begin{equation}
\label{QF}
  \begin{split}
    Q_2^F& = \Delta\\
   Q_4^F &= 2\Delta\\
   Q_6^F& = 16\Delta+ 8\Delta^3\\
   Q_8^F& =272\Delta + 416\Delta^3 + 32\Delta^5\\
   Q_{10}^F &= 7936\Delta + 24576\Delta^3 + 7680\Delta^5 + 128\Delta^7\\
   Q_{12}^F& = 353792\Delta + 1841152\Delta^3 + 1304832\Delta^5 + 128512\Delta^7 + 512\Delta^9.\\
  \end{split}
\end{equation}
We stress that eqs.  \eqref{QF} represent the constant terms entering
the r.h.s. of \eqref{GMitt}. The ferromagnetic eigenvalues of the
higher charges all vanish in our normalization.

\section{Numerical implementation}

\label{sec:num}

For our numerical calculations we considered the regime
$\Delta>1$. 
In this regime the XXZ chain is gapped: there are two
ground states which become degenerate in the $L\to\infty$ limit, but
there exists a finite gap between them and the next excited state. In
the limit of $\Delta\to\infty$ the Hamiltonian turns into the
classical Ising model, with its two ground states given by the N\'eel
and anti-N\'eel states. Therefore quenching to a large but finite
$\Delta$ can be considered a ``small quench'': we expect that all
physical quantities predicted by the GGE will be close to their
respective values
in the N\'eel state. Departure from these mean values should show up for
smaller $\Delta$. For our actual numerical calculations we chose
the values $\Delta=2,3,4,5$. 

We implemented the NLIE \eqref{NLIE1} 
and
solved it numerically by simple iteration, which
converged even when the higher charges were added. It is known
that the iteration method becomes ineffective in the low-temperature
regime, where $\beta_2\Delta$ is large. In this case
a different formulation of the NLIE can be set up
\cite{kluemper-review}; however this was not needed for our purposes.

We also implemented the factorized formulas of
\cite{XXZ-finite-T-factorization} for the short-range correlators
\begin{equation*}
  \vev{\sigma^a_1\sigma^a_2},\  \vev{\sigma^a_1\sigma^a_3} 
\text{ and }  \vev{\sigma^a_1\sigma^a_4} 
\end{equation*}
for $a=z,x$. We checked our programs by calculating these correlators
in the purely thermal case and comparing them to the values published
in \cite{XXZ-massive-corr-numerics-Goehmann-Kluemper} and we found
complete agreement.

We tested our numerics for the higher conserved charges by evaluating
them in the purely thermal case and at low temperatures ($\beta_j=0$
for $j>3$ and $\beta_2\to\infty$). In this limit the
thermal averages quickly approach the eigenvalues in the two ground states,
because the Hamiltonian is gapped. On the other hand, the ground state
mean values can be found by standard Bethe Ansatz calculations \cite{Takahashi-book}:
\begin{equation}
\label{QG}
\lim_{\beta_2\to\infty}  \vev{Q_{2j}}=-
\sum_{n=-\infty}^\infty (2n)^{2(j-1)}\frac{e^{-\eta |n|}}{\cosh(\eta n)}.
\end{equation}
We compared our numerical results with the formula above and found
convincing agreement. Note that the $\Delta\to\infty$ limit of \eqref{QG}
reproduces the leading terms of \eqref{NN}, given that the
normalization \eqref{GMitt} is used. This is a consequence
of the fact that the states $(\ket{N}\pm \ket{AN})/\sqrt{2}$
become the two ground states in the
$\Delta\to\infty$ limit.

As a further numerical verification we also evaluated the high-temperature
limit of the mean values. At high-temperatures
any product of spin operators tends to zero, because the
Pauli matrices are traceless
operators. Therefore only the constant terms of $Q_{2j}$ contribute and
from \eqref{GMitt} we obtain
\begin{equation*}
  \lim_{\beta_2\to 0} \vev{Q_{2j}}=
-\frac{\bra{F}Q_{2j}^{\text{GM}}\ket{F}}{2 (\sinh\eta)^{2j-1}}.
\end{equation*}
We compared our numerical results to those calculated from \eqref{QF}
and found complete agreement\footnote{The high-temperature limit of
  $\vev{Q_{2j}}$ could be calculated also from \eqref{NLIEmderiv}-\eqref{Qj} using 
  $\lim_{\beta_2\to 0}\fa(\lambda)=1$.}. 

In order to find the numerical values of the Lagrange multipliers
$\beta_j^{(k)}$ we used the multi-dimensional Newton-Raphson
method\footnote{This was suggested by G\'abor Tak\'acs.}. The matrix elements of the Jacobian 
\begin{equation*}
 \frac{\partial\vev{Q_m}}{\partial\beta_n}
\end{equation*}
are easily obtained by taking a further derivative of \eqref{Qj}. The
Newton-Raphson method converged if started from a purely thermal state
with an appropriate $\beta_2$.

We observed that the numerics became sensitive to the choice of
the integration contour for the NLIE as we add the higher
charges. There are always two numbers $0<\alpha_1$ and $\alpha_2<\eta/2$
such that every $\alpha$ chosen from the interval
$[\alpha_1,\alpha_2]$ gives the same answer for all quantities up to a
desired numerical accuracy. However, the
available interval for $\alpha$ shrinks with the growing number of the
charges. This can be explained by the fact that the higher
source terms for the NLIE are more and more singular, which in turn  requires that
the integration contour is far enough from the origin in order for the calculations
to be stable. As a rule of thumb we used the values $\alpha=0.95\times \eta/2$.

In order to gather information about the Bethe roots of the QTM 
 we calculated the function
$L(\lambda)=(1+\fa(\lambda))$ along the imaginary axis using \eqref{NLIE1}. 
Zeroes of $L(\lambda)$ determine the positions of the roots, which
 are known to be purely imaginary in the thermal case (at zero
 magnetic field). 
In all our examples it was found that although the position of the
roots changes considerably as compared to the thermal case, they always stay on
the imaginary axis. This provides a strong justification for the
validity of the NLIE \eqref{NLIE1}. 

Having convinced ourselves that our numerics is stable, 
 we
computed the predictions for the short-range correlators. 
Our calculations were carried out with a total of $n=400$
discretization points. When $\alpha$ was chosen from the stability
interval, then the numerical values did not change up to 6 digits when
we increased $n$ to 600, or when we slightly changed $\alpha$.
We conclude that the results presented below are accurate 
up to the last digit.

Tables \ref{CD1} and \ref{CD2} include our numerical results. The
correlation functions are also plotted as a function of the truncation
level in figures
\ref{fig1} and \ref{fig2}.

\bigskip

Regarding the Lagrange multipliers our data is not sufficient to
determine whether they are convergent
as a function of the truncation level.
Note that convergence of the $\beta_j$ themselves is not required by
physical principles, because they  can not be measured. In fact higher
charges can modify the lower $\beta_j$ considerably, because they 
involve terms which are already present in the lower charges. 
We also calculated the cumulated coefficients
$\gamma_1$ and $\gamma_2$ defined as
\begin{equation*}
 - \sum_{l=2}^k \beta_l^{(k)} Q_l=
-\gamma_1 \sum_j \sigma_{j}^z\sigma_{j+1}^z-\gamma_2\sum_j \sigma_{j}^z\sigma_{j+2}^z+\dots
\end{equation*}
Our results at $\Delta=5$ are below:
\begin{center}
\begin{tabular}{|c||c|c|c|c|c|c|}
\hline
$k$ & 2 & 4 & 6 & 8 & 10 & 12 \\
\hline
$\gamma_1$ &19.462 &  18.632 &  20.264 &  20.108  & 20.793 &  20.774\\
\hline
$\gamma_2$&  0 &  0.23873  & 0.22747 &  0.28444&   0.28171 &  0.30213\\
\hline
\end{tabular}
\end{center}
These data are not sufficient to determine whether $\gamma_1$ and $\gamma_2$ are convergent.

\bigskip

Regarding the correlators we observe the following:
\begin{itemize}
\item For all $\Delta$ and all distances the $z-z$ correlators increase, the $x-x$
  correlators decrease as a function of the truncation level. The
  physical reason for
  this is evident: The N\'eel state has maximal
  $z-z$ (anti-)correlations and identically zero $x-x$
  correlations. Addition of higher conserved charges should reproduce
  more details of the initial state, leading to the observed
  behavior.
\item All correlators seem to converge for $\Delta=4$ and
  $\Delta=5$. In order to convince ourselves we
  plotted the differences between the predictions at truncation level
  $k$ and $k+2$ on a log-scale. Fig. \ref{Hi2} (a) shows the
  change of $\vev{\sigma_1^z\sigma_2^z}$. Apart from a cycle with
  period 2 (which is a peculiarity of the N\'eel state, see below) we
  clearly see the exponential decay of the differences. 
Fig. \ref{Hi2} (b) shows the changes of
  $\vev{\sigma_1^x\sigma_3^x}$. Here we see the same behavior with
  approximately the
  same exponent, except for the points at $k=2$. The fact that the
  truncated GGE predictions at $k=2$ do not fit the exponential decay
  is not surprising: it is expected that for local operators spanning
  $j$ sites the first $j$ charges need to be added to reach the GGE
  prediction or at least the region where exponential convergence sets
  in. Similar behaviour was observed in the Ising chain in \cite{essler-truncated-gge}.
\item At $\Delta=2$ and $\Delta=3$ the numerical data does not show
convergence in the regime $k\le 12$. In fact, plotting the differences on a
  log scale as before results in seemingly random data
  (Figs. \ref{Hi2} (c) and (d)). The correlations are monotonic
  functions of
  $k$ and they are bounded so eventually they have
  to converge. We conjecture that there is exponential
  convergence also for  $\Delta=2$ and $\Delta=3$, but it starts at some $k=k_c$ with
  $k_c>12$.
\item For large $\Delta$ all GGE predictions are close to the N\'eel values,
where the $z-z$ correlations are maximal and the $x-x$ correlations
are zero. For smaller $\Delta$ we see gradual departure towards the
isotropic limit. 
\end{itemize}

Summarizing the above findings the following picture emerges. There
exists a ``convergence length'' $\kappa$ which depends on $\Delta$, such
that for \textit{every} short-range correlator $\ordo$
\begin{equation}
\vev{\ordo}^{(k)}=\vev{\ordo}_{GGE}+\left(\alpha+\beta(-1)^{k/2}\right) e^{-k/\kappa}+\dots,\text{
  for } k>k_c,
\label{GGEkonv}
\end{equation}
where $\alpha,\beta$ and $k_c$ depend of $\ordo$, and if $\ordo$ spans at
most $j$ sites then $k_c\ge j$. 
The convergence length is small when the pre-quench and post-quench Hamiltonian are
``close'', in our case when $\Delta$ is large.  
The dots on the r.h.s. of \eqref{GGEkonv}
denote contributions which decay faster than the leading
correction. The form of \eqref{GGEkonv} takes into account the
observed oscillation with period 2, which is a peculiarity of the
N\'eel state.

Due to the limitations of our data we do not perform any $k\to\infty$
fits of the correlators. For any practical purposes we suggest to take
the values at $k=12$. For larger $\Delta$ these data are already quite
accurate, whereas for smaller $\Delta$ they are to be understood as
lower or upper bounds, depending on the correlator in question. All
our data shows clear difference between the thermal prediction ($k=2$) and
the approximation of the full GGE ($k=12$). The differences are more
pronounced for the 3-site and 4-site correlators.

We conclude this section by providing a possible explanation for the
observed cycle of period 2 in Fig \ref{Hi2} (a) and (b). 
For any even $k$ the charge $Q_{k}$ has terms
proportional to $(\sigma_j^z\sigma_{j+m}^z-1)$ with $m=1\dots k/2$ \cite{GM-higher-conserved-XXZ},
where we already subtracted the ferromagnetic expectation value. In the N\'eel state these terms 
evaluate to $(-1)^{m}-1$; they vanish for every even $m$. 
Therefore, whenever $k=4m$, the last two-site $z-z$ operator
added to the GGE does not feel the antiferromagnetic
order of the N\'eel state, in contrast to the charges $k=4m+2$ which
are the ones who ``submit'' this information. This also
explains the peculiar behavior of the thermal multiplier
$\beta_2^{(k)}$:
its value changes considerably when a new charge with $k=4m+2$ is added, but it
almost stays the same when the truncation level is raised to
$k=4m+4$. 

\section{Conclusions}

We considered global quantum quenches in the XXZ spin chain and showed
that the Generalized Gibbs Ensemble can be implemented within the Quantum Transfer
Matrix framework, and it yields predictions for the long-time limit of
short-range correlators. The central idea was to construct
truncated GGE's with only a finite number of higher charges; we then
investigated the convergence of the predictions as the truncation
level is increased. 

In our  example we considered a quantum quench starting from the N\'eel
state. Quenching to a finite but large $\Delta$ we found that our
iterative procedure converges exponentially fast as a function of the
truncation level. On the other hand, for smaller $\Delta$
convergence was not yet reached in the truncated GGE's with the first
12 charges.

Correlations in the long-time limit are close to their values in the
N\'eel state when we quench to a large $\Delta$. On the other hand, quenching to
smaller $\Delta$ the system departs towards the isotropic
correlations.  An important result of the present
work is that the addition of the higher charges forces the
correlations towards their N\'eel values.
This is consistent with the general idea behind GGE, namely that the
existence of the higher charges constrains the dynamics and physical
observables can not relax to their thermal values.

\begin{figure}
  \centering
 \subfigure[$|\sigma^z_1\sigma^z_{2,3,4}|$ at
 $\Delta=5$]{\includegraphics[scale=0.7]{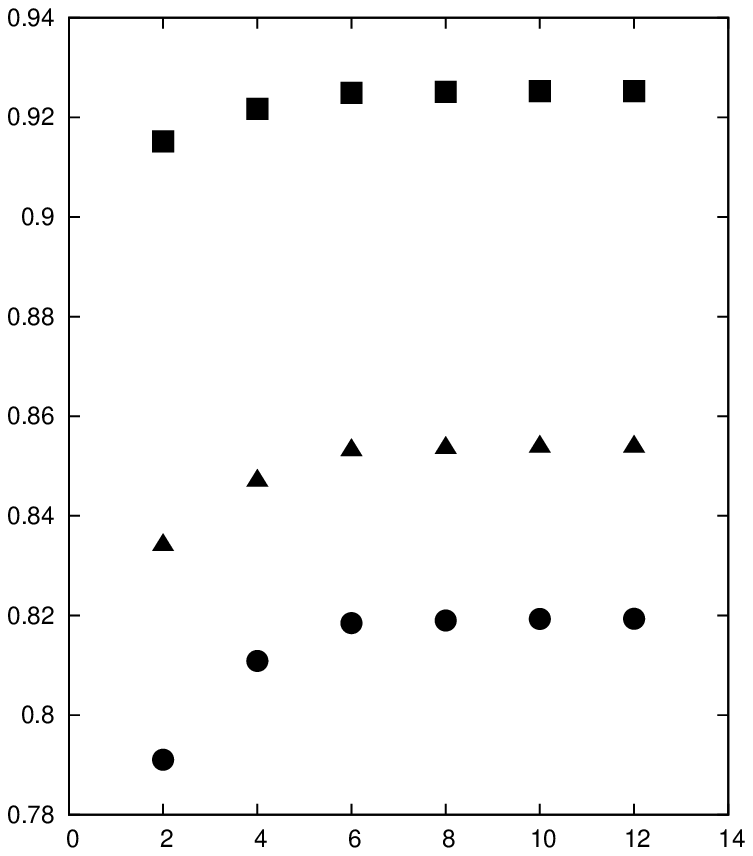}} 
\hspace{1cm}
 \subfigure[$|\sigma^x_1\sigma^x_{2,3,4}|$ at $\Delta=5$]{\includegraphics[scale=0.7]{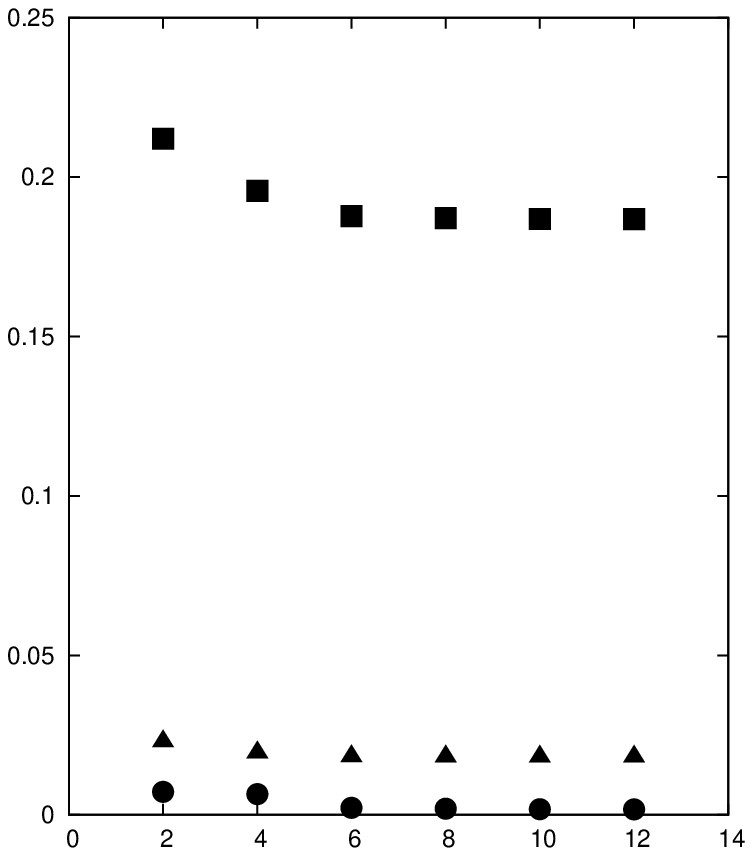}} 

 \subfigure[$|\sigma^z_1\sigma^z_{2,3,4}|$ at
 $\Delta=4$]{\includegraphics[scale=0.7]{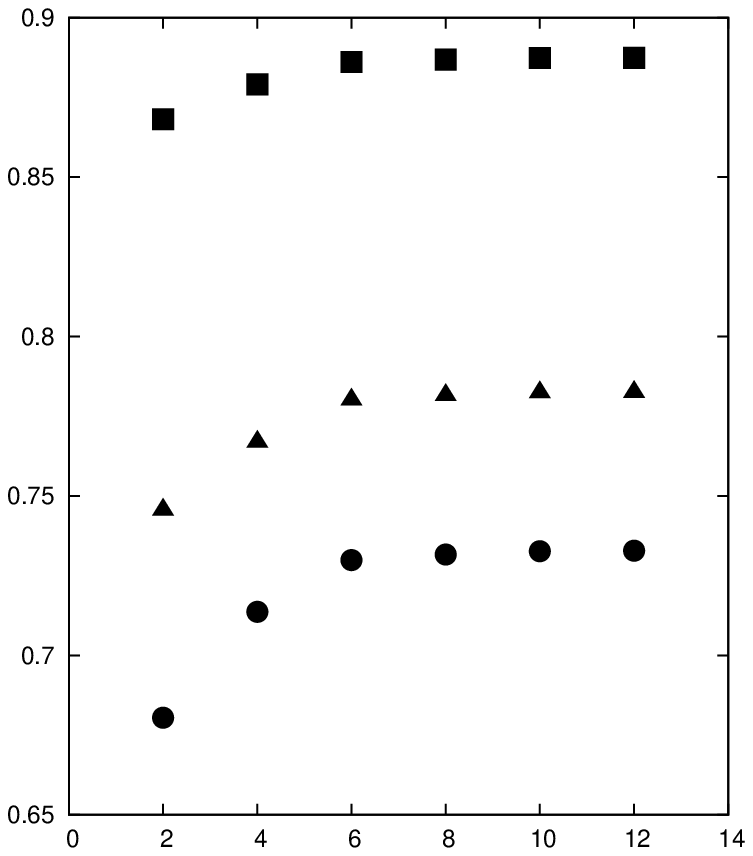}} 
\hspace{1cm}
 \subfigure[$|\sigma^x_1\sigma^x_{2,3,4}|$ at $\Delta=4$]{\includegraphics[scale=0.7]{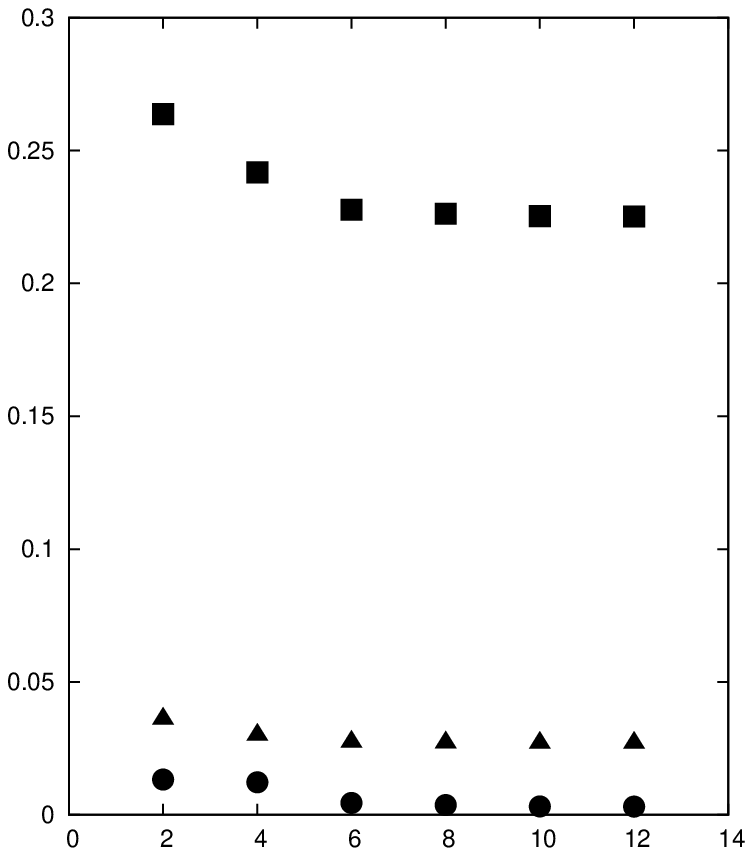}} 

 \subfigure[$|\sigma^z_1\sigma^z_{2,3,4}|$ at
 $\Delta=3$]{\includegraphics[scale=0.7]{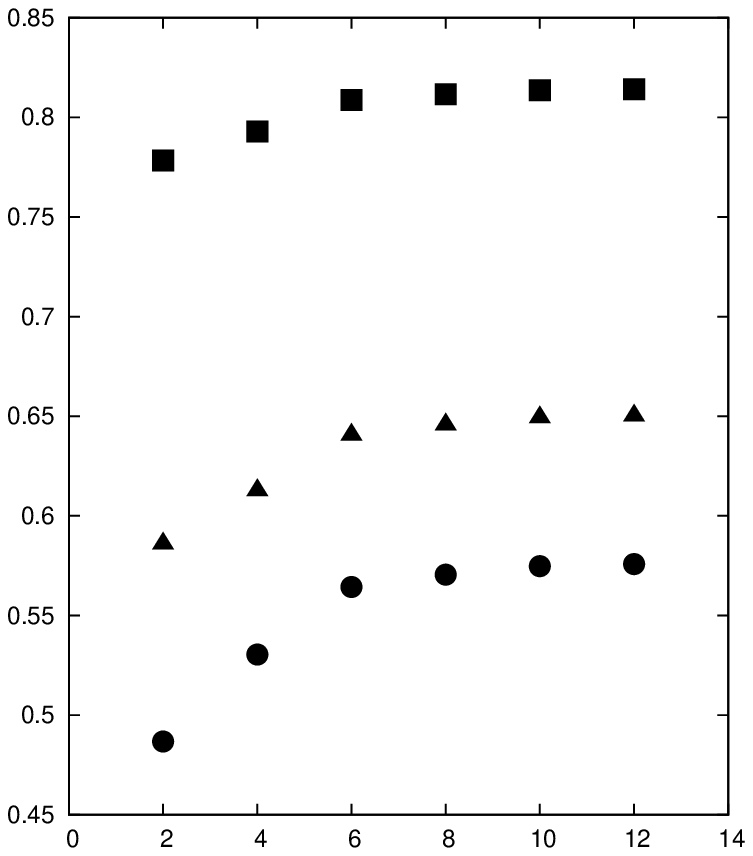}} 
\hspace{1cm}
 \subfigure[$|\sigma^x_1\sigma^x_{2,3,4}|$ at $\Delta=3$]{\includegraphics[scale=0.7]{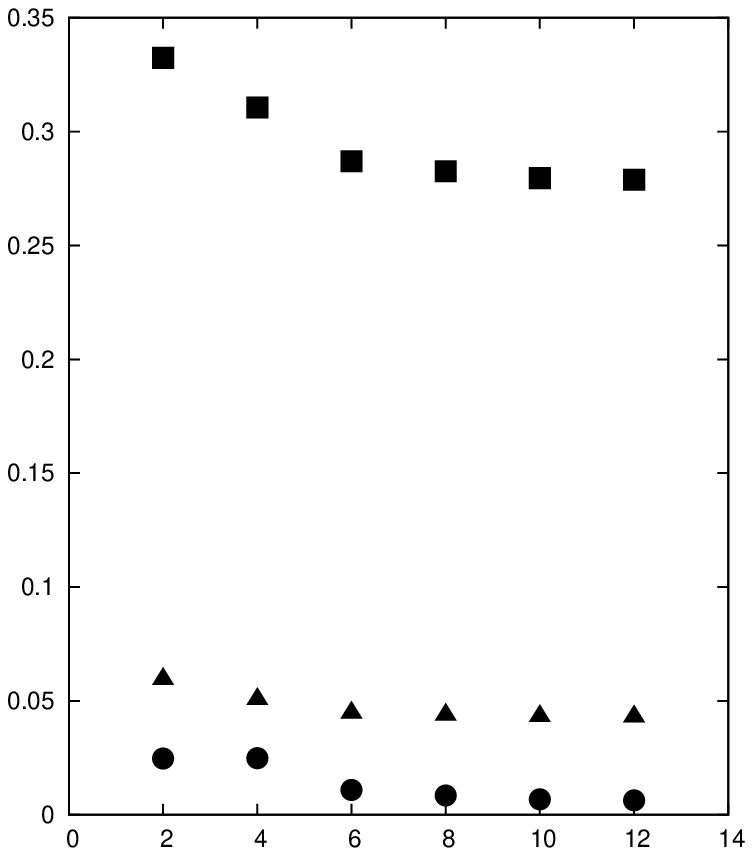}}
\caption{The short-range correlators predicted by the truncated GGE's
as a function of the truncation level. The top,
middle and bottom curves correspond to $\sigma^a_1\sigma^a_{1+j}$ with
$j=1,2$, and 3, respectively. We plotted the magnitude of the
correlators, the signs are given by $(-1)^j$.}
\label{fig1}
\end{figure}

\begin{figure}
  \centering
 \subfigure[$|\sigma^z_1\sigma^z_{2,3,4}|$ at $\Delta=2$]{\includegraphics[scale=0.7]{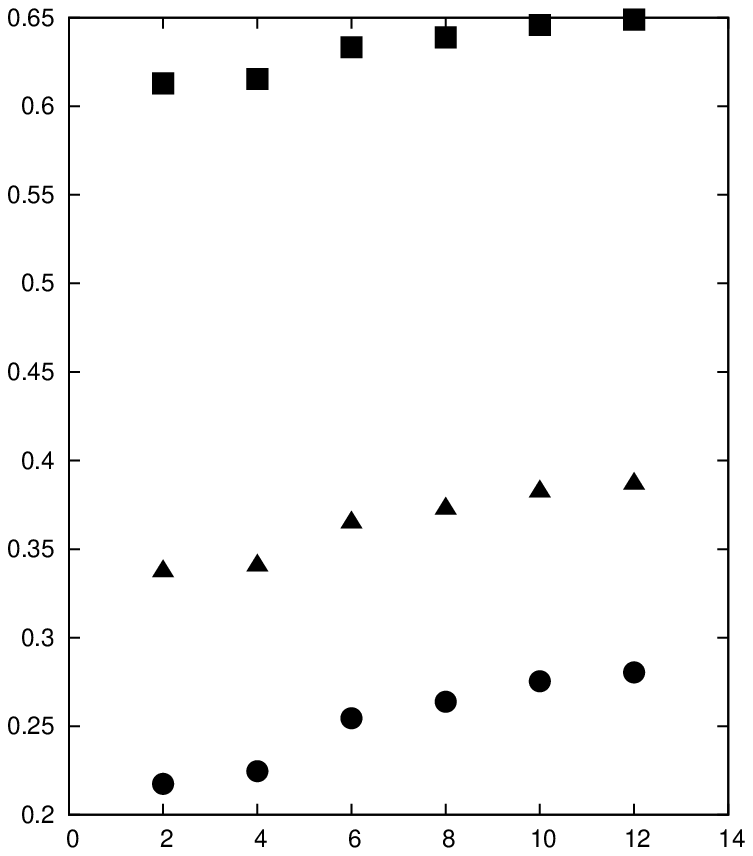}} 
\hspace{1cm}
 \subfigure[$|\sigma^x_1\sigma^x_{2,3,4}|$ at $\Delta=2$]{\includegraphics[scale=0.7]{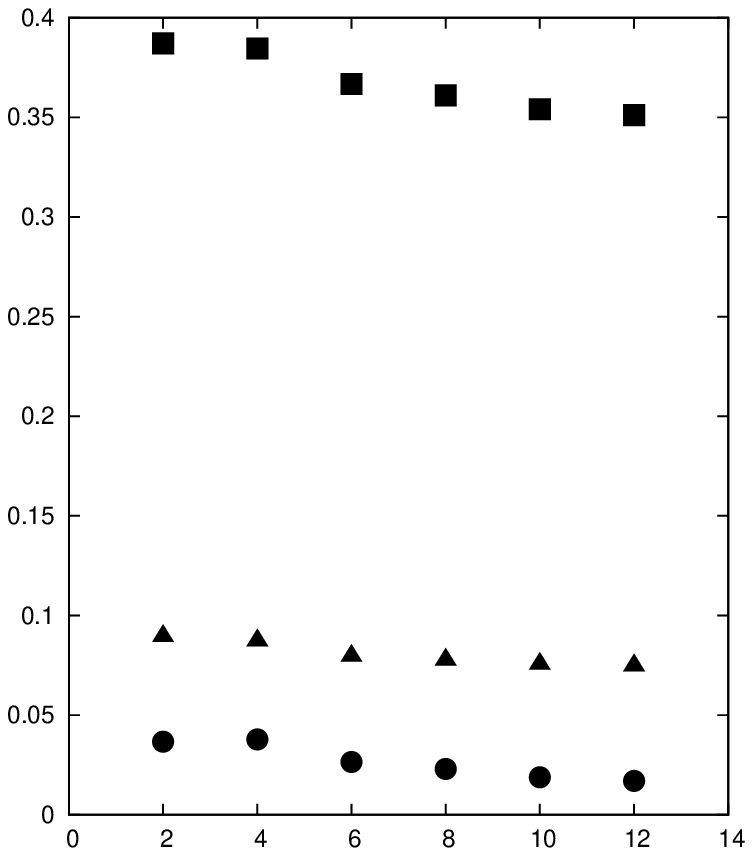}}
\caption{The short-range correlators predicted by the truncated GGE's
as a function of the truncation level. The top,
middle and bottom curves correspond to $\sigma^a_1\sigma^a_{1+j}$ with
$j=1,2$, and 3, respectively. We plotted the magnitude of the
correlators, the signs are given by $(-1)^j$.}
\label{fig2}
\end{figure}

\begin{figure}
  \centering
\psfrag{D4}{$\Delta=4$}
\psfrag{D5}{$\Delta=5$}
\psfrag{D2}{$\Delta=2$}
\psfrag{D3}{$\Delta=3$}
 \subfigure[$\ordo=\sigma_1^z\sigma_2^z$ for $\Delta=4$ (triangles)
 and $\Delta=5$ (squares)]{\includegraphics[scale=0.7]{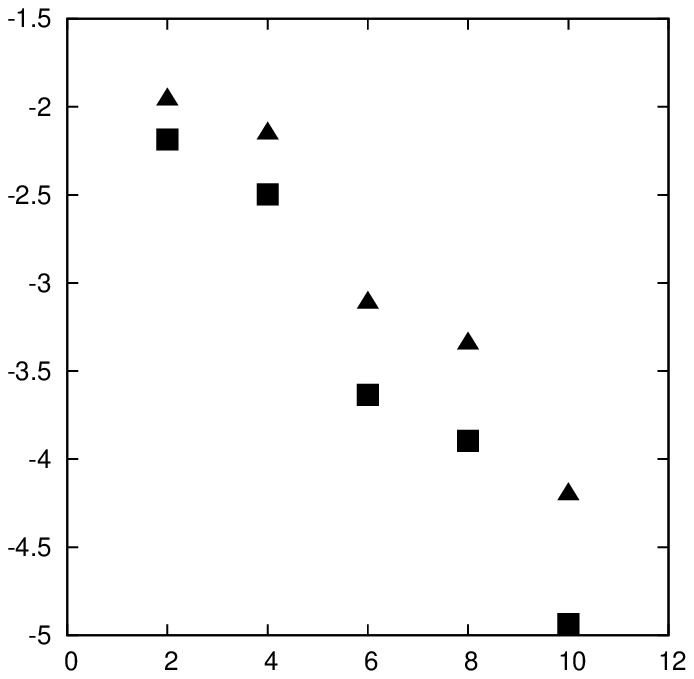}} 
\hspace{1cm}
 \subfigure[$\ordo=\sigma_1^x\sigma_3^x$ for $\Delta=4$ (triangles)
 and $\Delta=5$ (squares)]{\includegraphics[scale=0.7]{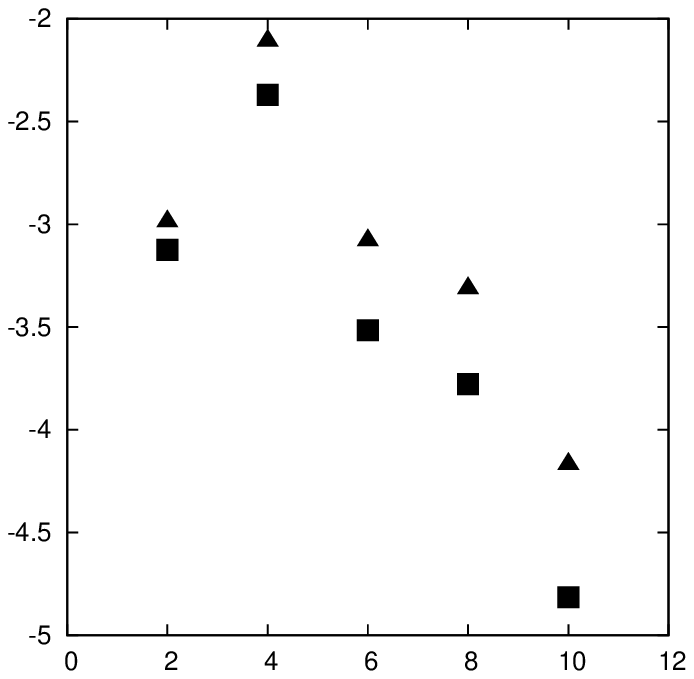}} 

 \subfigure[$\ordo=\sigma_1^z\sigma_2^z$ for $\Delta=2$ (triangles)
 and $\Delta=3$ (squares)]{\includegraphics[scale=0.7]{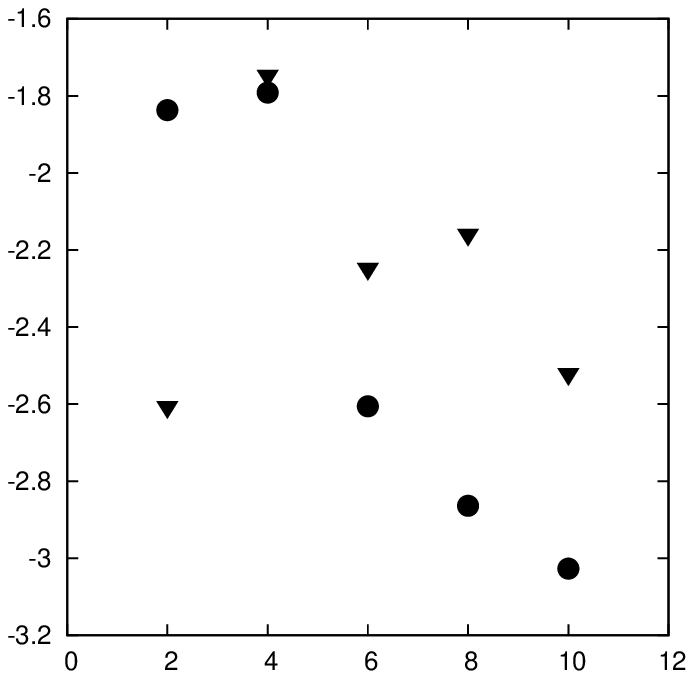}} 
\hspace{1cm}
 \subfigure[$\ordo=\sigma_1^x\sigma_3^x$ for $\Delta=2$ (triangles)
 and $\Delta=3$ (squares)]{\includegraphics[scale=0.7]{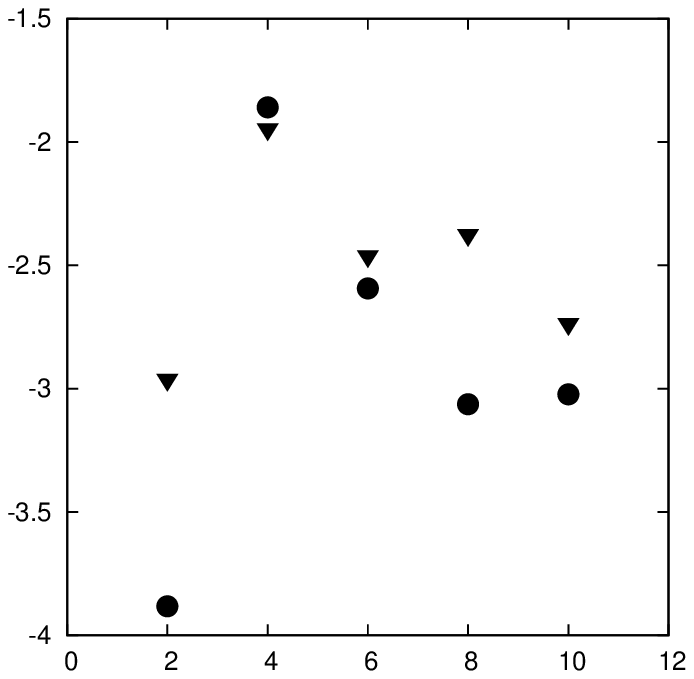}} 
\caption{ $\log(|\ordo^{(k+2)}-\ordo^{(k)}|)$ as a
  function of $k$.}
\label{Hi2}
\end{figure}

One of our original expectations was to find that adding $k$ charges to the GGE
would fix all correlators spanning at most $k$ sites, possibly with negligible
correction terms.  Our findings show that this is only true, when the
pre-quench and post-quench Hamiltonians are sufficiently close. Furthermore,
even in these cases the dependence on the truncation level looks
very similar for all correlators (see fig. \ref{fig1}). In our case we
found that the x-x correlations reveal the expected pattern:
exponential decay of the corrections sets in when the required number
of higher charges is already added (compare figs. \ref{Hi2}(a) and \ref{Hi2}(b)).

Our results show that the convergence length $\kappa$ (governing the
convergence as we add the higher 
charges) is
small for ``small quenches'' and that it is a new length scale 
which is independent of the correlation lenghts of the
system.
It is an interesting question how $\kappa$
 depends on the pre-quench and post-quench Hamiltonians in
more general situations. 

\bigskip

It would be interesting to consider the quench from the N\'eel state
at the isotropic point ($\Delta=1$).
Although this particular case is expected to exhibit the slowest
convergence as a function of the truncation level,
one advantage would be that in the XXX case all terms in the higher charges are
known analytically \cite{GM-higher-conserved-XXX}, therefore it might
be possible to reach much higher truncation levels.

The present methods could be applied to other quench situations as
well. The only requirement is that the mean values of the conserved
charges in the initial state should be evaluated exactly. In
principle this is possible even for a finite interaction quench
$\Delta\to\tilde\Delta$, because the charges of the post-quench Hamiltonian are
nothing more than certain combinations of Pauli matrices, and their
mean values in the pre-quench ground state could be obtained by
already available techniques.

\bigskip

We would like to stress that in this work we did not attempt to prove the GGE
hypothesis. Instead, our numerical results could be used as test of the
GGE. We determined both the thermal and GGE predictions, and we expect
that independent
numerical investigations could confirm one of the two, or they could point to a different
result. Numerical methods which simulate real-time dynamics typically involve a certain
amount of integrability breaking, therefore they are expected to
converge to our thermal 
predictions (the rows with $k=2$ in the Tables 1 and 2), possibly with
a pre-thermalization regime where the GGE applies.

\vspace{1cm}
{\bf Acknowledgments} 

\bigskip

We are grateful to G\'abor Tak\'acs and M\'arton Kormos for stimulating
discussions and useful comments on the manuscript. Also, we are
indebted to Jean-S\'ebastien Caux for many discussions about related
problems, which motivated us to study quench dynamics in XXZ chains.

\bigskip

\addcontentsline{toc}{section}{References}
\bibliography{../../pozsi-general}
\bibliographystyle{utphys}

\begin{table}
  \centering
  \begin{tabular}{|c||c|c|c|c|c|c|}
\hline
    $k$ & $\beta^{(k)}_2$ & $\beta^{(k)}_4$ &  $\beta^{(k)}_6$ &
    $\beta^{(k)}_8$ &  $\beta^{(k)}_{10}$ &  $\beta^{(k)}_{12}$ \\ 
\hline \hline
2& 1.425983 &  &  &  &  &   \\ \hline 
4& 1.446665 & -1.645590$\times 10^{-2}$ &  &  &  &   \\ \hline 
6& 1.809022 & -1.858076$\times 10^{-1}$ & 9.534796$\times 10^{-3}$ &  &  &   \\ \hline 
8& 1.856607 & -2.907996$\times 10^{-1}$ & 1.606714$\times 10^{-2}$ & -9.262052$\times 10^{-5}$ &  &   \\ \hline 
10& 2.058829 & -4.366422$\times 10^{-1}$ & 3.261496$\times 10^{-2}$ & -4.227379$\times 10^{-4}$ & 1.792828$\times 10^{-6}$ &   \\ \hline 
12& 2.084167 & -5.414107$\times 10^{-1}$ & 4.392467$\times 10^{-2}$ & -7.655830$\times 10^{-4}$ & 4.745685$\times 10^{-6}$ & -7.809887$\times 10^{-9}$  \\ \hline 
  \end{tabular}
\bigskip

  \begin{tabular}{|c||c|c|c||c|c|c|}
\hline
$k$ & $\vev{\sigma_1^z\sigma_2^z}$ & $\vev{\sigma_1^z\sigma_3^z}$ &
$\vev{\sigma_1^z\sigma_4^z}$ & $\vev{\sigma_1^x\sigma_2^x}$ &
$\vev{\sigma_1^x\sigma_3^x}$ & $\vev{\sigma_1^x\sigma_4^x}$  \\ \hline
\hline 
    2 & -0.612933 & 0.337184 & -0.217441 & -0.387067 & 0.089492 & -0.036689  \\ \hline
4 & -0.615407 & 0.340498 & -0.224563 & -0.384593 & 0.087323 & -0.037779  \\ \hline
6 & -0.633311 & 0.364773 & -0.254495 & -0.366689 & 0.079573 & -0.026466  \\ \hline
8 & -0.638963 & 0.372576 & -0.263839 & -0.361037 & 0.077668 & -0.023009  \\ \hline
10 & -0.645892 & 0.382295 & -0.275347 & -0.354108 & 0.075581 & -0.018792  \\ \hline
12 & -0.648906 & 0.386578 & -0.280383 & -0.351094 & 0.074730 & -0.016958  \\ \hline
  \end{tabular}

\bigskip

(a) $\Delta=2$

\vspace{2cm}

  \begin{tabular}{|c||c|c|c|c|c|c|}
\hline
    $k$ & $\beta^{(k)}_2$ & $\beta^{(k)}_4$ &  $\beta^{(k)}_6$ &
    $\beta^{(k)}_8$ &  $\beta^{(k)}_{10}$ &  $\beta^{(k)}_{12}$ \\ 
\hline \hline
2& 2.509646 &  &  &  &  &   \\ \hline 
4& 2.579696 & -2.307552$\times 10^{-1}$ &  &  &  &   \\ \hline 
6& 3.001455 & -6.465130$\times 10^{-1}$ & 4.852242$\times 10^{-2}$ &  &  &   \\ \hline 
8& 3.000383 & -8.891678$\times 10^{-1}$ & 7.641464$\times 10^{-2}$ & -7.568640$\times 10^{-4}$ &  &   \\ \hline 
10& 3.172756 & -1.182961 & 1.392459$\times 10^{-1}$ & -3.008985$\times 10^{-3}$ & 2.183419$\times 10^{-5}$ &   \\ \hline 
12& 3.173845 & -1.368100 & 1.759998$\times 10^{-1}$ & -4.936284$\times 10^{-3}$ & 5.064619$\times 10^{-5}$ & -1.331293$\times 10^{-7}$  \\ \hline 
  \end{tabular}
\bigskip

  \begin{tabular}{|c||c|c|c||c|c|c|}
\hline
$k$ & $\vev{\sigma_1^z\sigma_2^z}$ & $\vev{\sigma_1^z\sigma_3^z}$ &
$\vev{\sigma_1^z\sigma_4^z}$ & $\vev{\sigma_1^x\sigma_2^x}$ &
$\vev{\sigma_1^x\sigma_3^x}$ & $\vev{\sigma_1^x\sigma_4^x}$  \\ \hline
\hline 
2 & -0.778391 & 0.585901 & -0.486765 & -0.332413 & 0.059452 & -0.024717  \\ \hline
4 & -0.792940 & 0.612660 & -0.530385 & -0.310590 & 0.050655 & -0.024848  \\ \hline
6 & -0.808696 & 0.640589 & -0.564324 & -0.286956 & 0.044700 & -0.010937  \\ \hline
8 & -0.8115893 & 0.6456882 & -0.5704706 & -0.2826161 & 0.0437445 & -0.0084729  \\ \hline
10 & -0.8136196 & 0.6492727 & -0.5747833 & -0.2795706 & 0.0431133 & -0.0067519  \\ \hline
12 & -0.8140907 & 0.6501061 & -0.5757850 & -0.2788640 & 0.0429724 & -0.0063530  \\ \hline
  \end{tabular}

\bigskip

(b) $\Delta=3$

\vspace{1cm}

\caption{The Lagrange multipliers  of
  the truncated GGE's 
and the predictions for the short-range correlators 
as a function of the truncation level.
}
\label{CD1}
\end{table}

\begin{table}
  \centering
  \begin{tabular}{|c||c|c|c|c|c|c|}
\hline
    $k$ & $\beta^{(k)}_2$ & $\beta^{(k)}_4$ &  $\beta^{(k)}_6$ &
    $\beta^{(k)}_8$ &  $\beta^{(k)}_{10}$ &  $\beta^{(k)}_{12}$ \\ 
\hline \hline
2& 3.324318 &  &  &  &  &   \\ \hline 
4& 3.311865 & -4.272592$\times 10^{-1}$ &  &  &  &   \\ \hline 
6& 3.680208 & -1.055429 & 1.010927$\times 10^{-1}$ &  &  &   \\ \hline 
8& 3.670066 & -1.383075 & 1.524210$\times 10^{-1}$ & -1.791012$\times 10^{-3}$ &  &   \\ \hline 
10& 3.821611 & -1.908843 & 2.914045$\times 10^{-1}$ & -8.182288$\times 10^{-3}$ & 7.859512$\times 10^{-5}$ &   \\ \hline 
12& 3.822664 & -2.195990 & 3.660003$\times 10^{-1}$ & -1.294490$\times 10^{-2}$ & 1.682970$\times 10^{-4}$ & -5.214364$\times 10^{-7}$  \\ \hline 
  \end{tabular}
\bigskip

  \begin{tabular}{|c||c|c|c||c|c|c|}
\hline
$k$ & $\vev{\sigma_1^z\sigma_2^z}$ & $\vev{\sigma_1^z\sigma_3^z}$ &
$\vev{\sigma_1^z\sigma_4^z}$ & $\vev{\sigma_1^x\sigma_2^x}$ &
$\vev{\sigma_1^x\sigma_3^x}$ & $\vev{\sigma_1^x\sigma_4^x}$  \\ \hline
\hline 
2 & -0.868150 & 0.745564 & -0.680444 & -0.263701 & 0.036000 & -0.013281  \\ \hline
4 & -0.879100 & 0.766869 & -0.713671 & -0.241800 & 0.029953 & -0.012251  \\ \hline
6 & -0.8861283 & 0.7800751 & -0.7298895 & -0.2277434 & 0.0273182 & -0.0044624  \\ \hline
8 & -0.8868965 & 0.7815109 & -0.7316455 & -0.2262069 & 0.0270605 & -0.0036317  \\ \hline
10 & -0.8873471 & 0.7823539 & -0.7326757 & -0.2253059 & 0.0269172 & -0.0031453  \\ \hline
12 & -0.8874100 & 0.7824718 & -0.7328197 & -0.2251799 & 0.0268979 & -0.0030773  \\ \hline
  \end{tabular}
\bigskip

(a) $\Delta=4$

\vspace{2cm}

  \begin{tabular}{|c||c|c|c|c|c|c|}
\hline
    $k$ & $\beta^{(k)}_2$ & $\beta^{(k)}_4$ &  $\beta^{(k)}_6$ &
    $\beta^{(k)}_8$ &  $\beta^{(k)}_{10}$ &  $\beta^{(k)}_{12}$ \\ 
\hline \hline
2& 3.892437 &  &  &  &  &   \\ \hline 
4& 3.821985 & -5.729564$\times 10^{-1}$ &  &  &  &   \\ \hline 
6& 4.164352 & -1.447134 & 1.638562$\times 10^{-1}$ &  &  &   \\ \hline 
8& 4.156574 & -1.859024 & 2.415704$\times 10^{-1}$ & -3.060743$\times 10^{-3}$ &  &   \\ \hline 
10& 4.298095 &-2.782744 & 5.102938$\times 10^{-1}$ & -1.720161$\times 10^{-2}$ & 1.965757$\times 10^{-4}$ &   \\ \hline 
12& 4.299221 & -3.243448 & 6.472767$\times 10^{-1}$ & -2.681533$\times 10^{-2}$ & 4.010930$\times 10^{-4}$ & -1.338719$\times 10^{-6}$  \\ \hline 
  \end{tabular}
\bigskip

  \begin{tabular}{|c||c|c|c||c|c|c|}
\hline
$k$ & $\vev{\sigma_1^z\sigma_2^z}$ & $\vev{\sigma_1^z\sigma_3^z}$ &
$\vev{\sigma_1^z\sigma_4^z}$ & $\vev{\sigma_1^x\sigma_2^x}$ &
$\vev{\sigma_1^x\sigma_3^x}$ & $\vev{\sigma_1^x\sigma_4^x}$  \\ \hline
\hline 
2 & -0.9151778 & 0.8340633 & -0.7910349 & -0.2120556 & 0.0229398 & -0.0072106  \\ \hline
4 & -0.9217059 & 0.8469762 & -0.8108822 & -0.1957353 & 0.0194593 & -0.0064598  \\ \hline
6 & -0.9248945 & 0.8531096 & -0.8184590 & -0.1877637 & 0.0182649 & -0.0021925  \\ \hline
8 & -0.9251263 & 0.8535539 & -0.8190064 & -0.1871842 & 0.0181865 & -0.0018872  \\ \hline
10 & -0.9252532 & 0.8537971 & -0.8193061 & -0.1868671 & 0.0181458 & -0.0017203  \\ \hline
12 & -0.9252648 & 0.8538195 & -0.8193336 & -0.1868380 & 0.0181422 & -0.0017050  \\ \hline
  \end{tabular}

\bigskip

(b) $\Delta=5$

\vspace{1cm}

\caption{The Lagrange multipliers  of
  the truncated GGE's 
and the predictions for the short-range correlators 
as a function of the truncation level}
\label{CD2}
\end{table}

\end{document}